# Intentional and serendipitous diffusion of ideas: Evidence from academic conferences

Misha Teplitskiy[1][*][†], Soya Park[2][†], Neil Thompson[2,3], David Karger[2]

1. *University of Michigan School of Information*
2. *MIT Computer Science and Artificial Intelligence Laboratory*
3. *MIT Initiative on the Digital Economy*

This paper investigates the effects of seeing ideas presented in-person, even when they are easily accessible online. Presentations may increase the diffusion of ideas intentionally (when one attends the presentation of an idea of interest) and serendipitously (when one sees other ideas presented in the same session). We measure these effects in the context of 25 computer science conferences using data from the scheduling application *Confer*, which lets users browse papers, *Like* those of interest, and receive schedules of their presentations. We address endogeneity concerns in presentation attendance by exploiting scheduling conflicts: when a user *Likes* multiple papers that are presented at the same time, she cannot see them both, potentially affecting their diffusion. Estimates show that being able to see presentations increases citing of *Liked* papers within two years by 1.5 percentage points (62.5% boost over the baseline rate). Attention to *Liked* papers also spills over to non-*Liked* papers in the same session, increasing their citing by 0.5 percentage points (125% boost), and this serendipitous diffusion represents 30.5% of the total effect. Both diffusion types were concentrated among papers semantically close to an attendee's prior work, suggesting that there are inefficiencies in finding related research that conferences help overcome. Overall, even when ideas are accessible online, in-person presentations substantially increase diffusion, much of it serendipitous.

† These authors contributed equally.
\* To whom correspondence should be addressed: Misha Teplitskiy, tepl@umich.edu

# 1. Introduction

Innovators in science and technology often build on the ideas of others by recombining them in novel ways (Fleming, 2001; Uzzi et al., 2013; Weitzman, 1998). Consequently, to understand the rate and direction of innovation it is crucial to understand how innovators encounter and learn others' ideas, that is how ideas diffuse. Yet idea diffusion is notoriously difficult to study. As Paul Krugman famously quipped, "knowledge flows … are invisible; they leave no paper trail by which they may be measured and tracked" (quoted in (Jaffe et al., 1993)). Consequently, scholars have had to make do with indirect or aggregated measures, and key questions about the drivers of diffusion remain open.

In particular, a persistent finding in the voluminous literature on diffusion is that its amount and intensity are associated with geographical proximity (Azoulay et al., 2011; Belenzon & Schankerman, 2013; Duede et al., 2024; Jaffe et al., 1993; Kabo et al., 2014; Lin et al., 2023; van der Wouden & Youn, 2023; Wuestman et al., 2019). The association may be caused by a variety of factors (Boschma, 2005) and a robust debate asks whether the primary factor is interpersonal ties between innovators (Head et al., 2019), the geographical concentration of related ideas (Wuestman et al., 2019), or something else. Scholars also debate at what scale the association is meaningful, if at all (Thompson & Fox-Kean, 2005), and even if it was meaningful in the past, whether recent technological progress supporting remote communication has made it obsolete (Head et al., 2019; Presidente & Frey, 2022).

A challenge to progress in these debates is that most studies examine data on the outputs of teams, such as patents and research papers, making it difficult to establish processes at the individual-level and that lead up to those outputs. In particular, it is plausible that the effects of geographical proximity on diffusion are driven by the benefits of in-person communication (Atkin et al., 2022; Azoulay et al., 2011; Torre, 2008) relative to online or another modality. The importance of the online modality has only increased with the COVID pandemic and its long-term effects on technology and the nature of work (Barrero et al., 2023) and a burgeoning literature seeks to understand the multifaceted consequences. Exemplary studies in this area typically randomize individuals to collaborate on and learn ideas *either* online vs. offline (Brucks & Levav, 2022; Kofoed et al., 2021). However, increasingly often individuals face an information environment where the same ideas are available online *and* in-person. It is possible that when online access to ideas exists seeing them communicated in-person adds little if any value. This paper contributes to the debate on geographical proximity and diffusion by assessing the effect of in-person communication of ideas when they are easily accessible online.

To do so we use the setting of academic conferences. Conferences are ubiquitous in scientific research and many other industries, and their future has been the subject of significant debate (Jarvis et al., 2021; Klöwer et al., 2020; Reshef et al., 2020; Thorp, 2020). Besides their intrinsic importance as a method of organizing and communicating (T. T. Hansen & Budtz Pedersen, 2018;

Hauss, 2020), conferences provide an attractive setting to study the value-add of in-person communication. In prior decades, researchers looking to read the newest papers could only find them once they were published, via personal connections, or as presentations at conferences and seminars. In that information environment, the choice to attend a conference presentation represented a choice between accessing the information or not. With the growing accessibility of papers online, the choice to attend presentations represents a choice between seeing information presented in-person and accessing it online vs. only accessing it online.

Measuring the effect of in-person presentations on diffusion is challenging. First, accumulating evidence of idea diffusion takes time. A researcher may be influenced by a conference presentation that inspires a paper published many years later. Second, researchers do not choose to attend paper presentations at random. This means that the natural comparison group for presentations attended – those not attended - will be on average less relevant or of lower quality. Consequently, when observational studies find that attendees self-report learning important information at presentations, *e.g.* (Lalonde et al., 2007), it is unclear if that outcome would have occurred even without the presentation.

We address these challenges with unique data from conferences and a novel identification strategy based on scheduling conflicts. Sometimes, a conference attendee finds that two or more presentations of interest are scheduled at the same time. When these scheduling conflicts occur, the person is on average less able to see either presentation in-person. In this way scheduling conflicts affect attendees' opportunities to attend presentations, without affecting their online access to the papers. We highlight that this variation is in the *opportunities* to attend presentations rather than actual attendance, and our research design can be interpreted as yielding intention-to-treat estimates of actual attendance. Thus, if we could instead measure actual attendance, we would expect the effects to be even higher than those we report here.

Our main identifying assumption is that, conditional on fixed effects for the user and the presented paper, a person's scheduling conflicts are plausibly random. In other words, we exploit the variation that a particular person may have a scheduling conflict with a particular paper but another person may not. User and paper fixed effects enable us to avoid confounding the results by a paper's quality, topic, or position in the overall schedule, or the user's propensity to *Like,* attend, or cite papers.

We apply this strategy to unique data from the conference scheduling application *Confer*, which lets attendees screen ("*Like*") papers and generates a personalized schedule of their presentations. The application was deployed at 25 computer science conferences between 2013-2022, and our analytic sample includes 5996 papers, 2481 users, and 83,726 *Likes*. We then identify users of this application in the bibliometric database OpenAlex and track whether they cited the presented papers in their subsequent work. We find that even when individuals have access to papers online, being able to see them in person increases citations within two years by 65%.

Conferences are also an attractive setting to study serendipity. The key role that serendipity has played in the history of many innovations is well documented (Balzano, 2022; Busch, 2022; Copeland et al., 2023; Merton & Barber, 2011; Simonton, 2004; Yaqub, 2018). Evidence is also accumulating that serendipity can to some extent be engineered by exposing innovators to unsought ideas through policy or other exogenous events (Catalini, 2017; Lane et al., 2021; Nahm et al., 2023). There is much less clarity on how big a role serendipity plays in the normal, everyday conduct of research. Are the colorful historical anecdotes of serendipity swamped by equally colorful but intentional events? And which serendipitous ideas are likeliest to be taken up?

Studying conferences enables us to contribute to the literature on serendipity by, first, providing a quantitative estimate of its role in diffusion. Many academic conferences present attendees with several tracks and many papers to choose from. Attendees will generally screen papers to select a few of interest, possibly attend those presentations, and utilize some of them in their subsequent work. We refer to this type of diffusion as *intentional.* Attendees may also encounter papers incidentally. A common type of incidental exposure is when an individual attends a session with a paper of interest and because of decorum or other pragmatic reasons remains in the room for other paper presentations, which she may have earlier screened out or not been aware of. If these unsought papers are later utilized we refer to this as *serendipitous* diffusion.

Following the intuition that attendees are likely to see multiple papers in a session, we focus on the papers a user did not *Like* but that are presented in the same sessions as *Liked* papers. We find that when attendees could see sessions with *Liked* papers, they were 125% more likely to cite the non-*Liked* papers. A back-of-the-envelope calculation suggests that this serendipitous diffusion represents 30.5% of the total citations induced by presentations. This estimate shows that even when serendipity is operationalized as one arguably narrow type, it represents a very substantial fraction of the diffusion.

Lastly, examining treatment effect heterogeneity illuminates which ideas are likeliest to diffuse through presentations. In particular, we measure the semantic distance between a conference attendee's papers before the conference and papers they intend to learn about or encounter serendipitously. We find that in both cases the effects are concentrated on papers semantically close to the attendee, with the effect for distant papers statistically indistinguishable from zero. That both intentional and serendipitous effects from prsentations are so important, even for work quite close to an attendees own research, corroborates existing work showing that information frictions in science are substantial (Azoulay et al., 2019; D. Feenberg et al., 2016). Furthermore, it suggests that efforts to exogenously expose scientists to semantically distant ideas, for example via interdisciplinary events, may struggle to achieve diffusion.

Overall, the study contributes to the debate on what drives the association between geographical proximity and diffusion by examining diffusion in an exceedingly common but rarely studied information environment – where the same ideas are available online and via an in-person modality, in this case presentations. The results show that in-person presentations greatly increase diffusion above the online-access baseline, both by increasing the uptake of ideas one intends to learn about and encounters serendipitously.

The rest of this paper is organized as follows. The next section reviews the literature on conferences and states the research questions. Section 3 describes the data and elaborates the identification strategy. Section 4 presents the three sets of results on intention diffusion, serendipitous diffusion, and heterogeneity across semantic distance. Section 5 concludes with limitations and implications.

## 2. Diffusion through presentations

In this section we review the several mechanisms through which conference presentations may drive diffusion of ideas beyond the level resulting from other communication modalities, in particular online access. We call these mechanisms awareness, engagement, and maturation. Furthermore, we distinguish between ideas that pass an initial screening, or do not pass or were never screened, and call their diffusion intentional or serendipitous, respectively. Lastly, we highlight the importance for diffusion of the semantic distance between an idea and an individual's existing research. Following existing work, we operationalize the diffusion of academic ideas using citations to papers (Azoulay et al., 2011; Leon & McQuillin, 2018).

Meetings and conferences play a substantial role in science and technology and a growing literature attempts to identify their many functions and consequences (Atkin et al., 2022; T. T. Hansen & Budtz Pedersen, 2018; Hauss, 2020). For example, studies have examined how conferences help attendees network (Campos et al., 2018; Chai & Freeman, 2019; Lane et al., 2021) and crystallize and legitimize new fields (Garud, 2008; Gross & Fleming, 2012). But arguably the most important function of academic conferences is diffusing ideas through presentations. Papers can diffuse through many other channels; for instance, a study of a random sample of citations found that most originated from researchers searching databases like Google Scholar (Teplitskiy et al., 2022). Researchers can and do cite papers without ever seeing them presented, leading to a baseline citation rate.

How do conference presentations increase citing above the baseline rate? The mechanisms will vary with whether papers are screened-in or not. Individuals generally do not choose to attend presentations at random, but first screen them, for example by reading an abstract, and attend those that are screened-in. Consequently, individuals will have at least superficial knowledge, *i.e.* awareness, of screened-in papers. Presentations may increase citing of screened-in papers by, first, improving engagement. A person may encounter an interesting abstract at a conference or elsewhere and plan to read the paper later but fail to execute the plan (Buehler et al., 1994). In-person conference presentations help execute the plan by providing a commitment mechanism (Bryan et al., 2010): the decision to attend a conference in the first place, often made long in advance of it, largely commits the individual to go, the concreteness and immediacy of the presentation's schedule may provide additional commitment (Trope & Liberman, 2003), and once an individual attends a session, it may be difficult to leave in the middle because of decorum or other pragmatic reasons. The link between in-person communication and improved engagement with ideas is further suggested by studies of schools (Jack et al., 2023) and universities, where students have described trouble concentrating when ideas are communicated online (Kofoed et al., 2021).

Second, academic presentations often provide opportunities for attendees to give feedback to the presenter (Rose et al., 2022). The feedback may improve the ideas and make them more attractive for citing. This mechanism has been called *maturation* (Leon & McQuillin, 2018). In our setting, the maturation mechanism is ruled out because the conferences comprising the data were archival, *i.e.* not works-in-progress.

We refer to an attendee's citing of screened-in papers as "intentional diffusion" since the attendee expressed explicit intent to learn about them, and ask

*RQ1*: *Do in-person presentations increase intentional diffusion?*

For ideas that a conference attendee initially screens out or never gets to screen, presentations may act via the mechanisms described above and the additional mechanism of *awareness*. Studies of scientific knowledge flows find that due to the overabundance of papers, researchers are often not aware of even the relevant ones (Azoulay et al., 2019; D. R. Feenberg et al., 2015). In-person presentations may increase awareness through incidental exposure. Papers are often organized into sessions, and it can be difficult to leave a session immediately before or after seeing the presentation(s) of interest. Consequently, attendees' attention to presentations of screened-in papers may spill over to other presentations in the same session. We refer to the citing of these other ideas as "serendipitous diffusion," since it does not fall into the intentional category, and ask:

*RQ2*: *Do in-person presentations increase serendipitous diffusion?*

A key question regarding serendipitous diffusion is its magnitude relative to intentional diffusion. Many of humanity's most prominent discoveries – from Velcro to Viagra – can be attributed to serendipity (Busch, 2020; Meyers, 2007). The literature has also established that serendipity can be engineered top-down through policy or other exogenous events. For example, rearranging the locations of where individuals work (Catalini, 2017; Lee, 2019) and access information (Nahm et al., 2023), or assigning them at random to different conference rooms (Lane et al., 2021), can all lead to diffusion and collaborations that would not have occurred otherwise. However, such exogenous changes are relatively rare. How much serendipity is there in the normal course of research? Do the historical anecdotes and exogenous changes play a minor role in research relative to intentional diffusion and discoveries?

Besides the intellectual interest, the question of magnitude has policy implications. Different communication modalities afford different levels of serendipity: for instance, it may be easier to exit an online session of paper presentations without compromising decorum than an in-person one. If so, online sessions may generate less serendipitous diffusion and conference organizers need to consider whether this is an issue worth addressing. However, if serendipitous diffusion is minor even at in-person sessions, the implications for the organizers are also minor. Accordingly, we ask

*RQ3*: *What fraction of total diffusion is serendipitous?*

The effect of presentations on diffusion is likely to vary with the semantic distance between a paper's ideas and a person's research area. Previous work has found that the probability of citation between two papers is lower the more semantically distant they are (Baldi, 1998; Wuestman et al., 2019). On the one hand, presentations may increase a researcher's citing above this baseline particularly when a paper is distant, because she is less likely to encounter it outside of the conference. On the other hand, presentations may do little to increase the citing of distant papers for two reasons. First, presentations may increase awareness and engagement for these ideas but their utility, as measured by citations, may remain low. Second, our research design relies on variation in opportunities to see presentations, *i.e.* intent-to-treat effects of seeing presentations. Researchers may be less likely to take advantage of opportunities to see semantically distant presentations, *i.e.* comply less with the assignment to a high opportunity to see them.

This topic also has policy implications. It is often claimed that the organization of researchers into groups with clear boundaries, such as university departments, has the effect of creating information silos that slow down innovation (Jeppesen & Lakhani, 2010; Lifshitz-Assaf, 2018). If so, it is appealing for policy-makers and administrators to stimulate diffusion by exogenously exposing researchers to ideas outside of their silo, which they often find particularly influential (Duede et al., 2024). However, if presentations are ineffective at stimulating intentional or serendipitous diffusion of distant ideas, top-down efforts may prove equally ineffective.

*RQ4*: *How do presentations' effects on intentional and serendipitous diffusion vary across paper-person semantic distance?*

Existing research that addresses these questions convincingly is very limited. To our knowledge, the study that comes closest is by de Leon and McQuillin (2018), which estimates the causal effect of conferences on citations using the unexpected cancellation of a major political science conference. By comparing that year's conference to prior instances, as well as to a smaller annual conference used as a control, the authors estimate that being presented at a conference increased a paper's chances of being cited at least once within four years after the conference by five percentage points. However, the study left several questions unresolved. First, it could not identify which activities of the conference (*i.e.* presentations, banquets) caused the effects, since all activities were canceled. Second, the conferences were for works-in-progress, so many papers underwent substantial changes or were abandoned; the authors were able to track the citations of only 27% of the presented papers. Third, the setting is mechanically complicated by the possibility of maturation, *i.e.* feedback improving papers and citations. Lastly and perhaps most importantly, it is likely that attendees of those conferences did not have synchronous access to those ideas outside of the conferences. In contrast, our interest is in the effect of presentations in the increasingly common case where the same ideas are easily available online and as presentations.

# 3. Data and Methods

## 3.1. *Confer*

Many conferences use scheduling software deployed via apps and websites to help attendees create their schedules. We use data from one such platform, *Confer*, which lets a user browse papers and *Like* those that are of interest, and displays a convenient, individualized schedule[1]. *Supplementary Information (SI): 1. Illustration of the software* provides screenshots of the interface. Data from *Confer* comprise metadata on the presented papers, such as their titles and timeslots, the names and affiliations users used when registering, and users' *Liked* papers. *Confer* was deployed at 25 prominent computer science conferences held between 2013-2020, representing 9070 paper presentations and other events like banquets and tutorials. The raw data included 2,759 unique usernames and 105,442 *Likes* to paper presentations and other events like banquets and tutorials.

We excluded from analysis *Likes* that were to users' own papers, papers published before the conference (*i.e.* "Test of Time" award sessions), and non-research-paper events. Non-research-paper events were identified using a variety of heuristics, such as keywords like "coffee" appearing in the session name or there being more than eight works scheduled for presentation. *SI 3. Full vs. analytic sample* details how the 470 non-research-paper events were identified. After *Likes* to such events were excluded there remained 2,541 unique user names and 89,536 *Likes*.

### 3.1.1. Measuring learning intentions with *Likes*

*Liking* papers on *Confer* indicates which papers a user intends to learn about. For a particular user the non-*Liked* papers at a particular conference consist of ones screened-out and never screened. Some non-*Liked* papers will be presented in the same sessions as a user's *Liked* papers; the user is unlikely to learn these papers *intentionally*, but given the challenges of entering and leaving in-person sessions in the middle, may do so incidentally. We focus on these same-session papers to measure serendipitous diffusion, and for convenience refer to them as non-*Liked* papers, although technically they are only a subset of all non-*Liked* papers.

Users may also be incidentally exposed to papers outside of sessions with *Liked* papers, although our research design gives us little leverage for such cases. We highlight that measuring the amount of serendipitous diffusion only through non-*Liked* papers in certain sessions is conservative, as this represents only one type of incidental exposure. The number of non-*Liked* user-paper pairs in the analytic sample is 108,402. The total number of user-paper pairs in the analytic sample was 192,128.

---

[1] Confer has been recently discontinued. An archived version is available at https://web.archive.org/web/20221222092038/https://confer.csail.mit.edu/. Accessed 2023-12-29.

## 3.2. Bibliometric data and matching

The *Confer* user and paper data was supplemented with bibliometric data from the databases *OpenAlex* (Priem et al., 2022)*, SemanticScholar* (Kinney et al., 2023), and *Crossref*. These databases are among the most comprehensive and widely used for bibliometric analyses.

### 3.2.1. Matching *Confer* users to *OpenAlex* records

Matching was done in two steps. First, we searched for a user's affiliation in *OpenAlex* to get its institution ID. We then performed an author search using the user's user name, institution ID (using the field *last_known_institution*), and the topic "Computer science" (*x_concepts* with ID C41008148). If this search yielded matches, we took the first match as the correct one[2]. If it yielded no matches, we repeated the search without using the institution constraint. This matching process failed to link seven users to OpenAlex author IDs. Author IDs with more than 500 papers were excluded because inspection revealed these to be disambiguation errors in *OpenAlex*. Inspection by one of the authors (M.T) of twenty randomly selected remaining user names showed that the matching was correct in 90% of the cases.

The three most common institutions of the users were University of Washington (86), Microsoft (68), and Carnegie Mellon University (68). The mean number of papers up to 2024 per user was 86.5 (*SD*=95.2, median=51).

### 3.2.2. Matching *Confer* papers to DOIs and metadata

Data from *Confer* included DOIs for 1591 papers. For the other papers and events (*e.g.* tutorials), we used a three-step matching process. First, we searched each paper in *SemanticScholar* using the title (with colons removed), the field (*fieldsOfStudy)* of "Computer Science," and the publication year equal to the year of the conference or the year prior. Second, if the search did not yield a result we repeated it after removing any part of the title after the colon. Third, for any remaining unmatched papers, we used the *Crossref* search API with a string consisting of "author names + paper title + conference year." We took the first result as the match. 9,664 papers were linked to DOIs in one of these ways, leaving 368 papers or events unmatched. To increase match quality, we used the DOI prefix (the digits before the slash), which identifies the publishing organization, *e.g.* ACM. For each of the 25 conferences, we identified the most common prefix, and excluded from the analytic sample DOIs with different prefixes. This exclusion reduced the number of included DOIs to 9231, and manual inspection did not uncover any inaccurate matches.

---

[2] Results are ranked by a relevance score, which is described in the API documentation https://docs.openalex.org/how-to-use-the-api/get-lists-of-entities/search-entities, accessed 2023-12-28. Relevance is calculated by text similarity and citations, with higher cited entities scoring higher.

## 3.3. Semantic distance and shared affiliation

Next, we obtained from *SemanticScholar* the papers' Specter2 embeddings, which are 768-dimensional vectors embedding the papers' titles and abstracts and yielding state-of-the-art performance on domain-specific tasks, such as predicting citations (Singh et al., 2022). We used these embeddings to calculate the semantic distance between a user's research and the presented paper as follows. We identified all of the user's papers published in the five years before the year of a particular paper's presentation and measured the average cosine distance between each of the user's papers and the presented paper. 13.1% of user-paper pairs had missing semantic distance because either the user had no papers before the conference or the embedding was not available in *SemanticScholar*.

Lastly, using *OpenAlex* we obtained the presented papers' author IDs, affiliations, and metadata on the papers citing them (up to 10,000). Using these metadata we identified if and when a user cited a paper after its presentation, as well as whether she and the authors shared an affiliation at the time of analysis in 2023. Data limitations prevented us from measuring shared affiliations dynamically.

## 3.4. Research design

### 3.4.1. Scheduling conflicts

The research design uses scheduling conflicts in a user's personal schedule for variation in her opportunities to see paper presentations. Our data did not include the exact scheduled time of each presentation but did have the session's starting time and presentations' order in the session. We assumed that presentations in sessions starting at the same time and with the same order were in the same ~15-minute timeslot. Scheduling conflicts are defined at the user-paper level: for example, if a user *Likes* only one paper in a timeslot, that user-paper pair has scheduling conflict level 0. If she likes two presentations in the same minute, both user-paper pairs have conflict level 1.

While defining scheduling conflicts for *Liked* papers is relatively straightforward, for serendipity papers it is less so. One reason is that users often *Like* multiple papers in a session, and each of them might have a different level of conflict. We take the average number of conflicts for all papers the user *Liked* in a session as the conflict level for the session's serendipity papers. For example, consider a user who *Liked* two papers in a session with four papers, one that is co-scheduled with one in another session and one that is co-scheduled with three in other sessions. Then for this user the two non-*Liked* (serendipity) papers in the session would have scheduling conflicts of (1+3)/2

= 2. For ease of exposition, we bin that variable as follows: "0" = [0,1), "1" = [1,2), "2" = [2,3), "3+" = [3, infinity).

### 3.4.2. Identification assumption

Scheduling conflicts vary the level or "dose" of treatment $\tau$ – opportunity to see presentations – that user-paper pairs receive. (This design can also be thought of as intention-to-treat, where the treatment is seeing presentations.) If a user *Likes* only one paper in a timeslot, then she has a full opportunity, or dose of $\tau$, to see it. If a user likes two papers in a timeslot, she can see only one of the presentations, so the average level of opportunity for these user-paper pairs is $\tau/2$. Similarly, for a timeslot with *n* papers, the average opportunity to see them is $\tau/n$.

A key concern for a causal interpretation of our results is that conference schedules are not exogenous, and so any user's scheduling conflicts may not be exogenous to their potential outcomes. To address the potential endogeneity we take two steps. First, we make only a conditional independence assumption, where we condition scheduling conflicts on paper and user fixed effects. Paper fixed effects control for potential underlying differences in papers' qualities, scheduling positions, presenter characteristics, and any other stable differences. This enables us to use only the variation in citing a paper that occurs when one user has many scheduling conflicts in seeing it while another does not. For example, if papers with more famous authors are scheduled in more attractive timeslots and attract more citations, such patterns will not affect our estimated treatment effect. Furthermore, we use user fixed effects to control for the overall propensity to attend presentations, publish and cite papers, and any other stable differences.

Second, we test for any possible remaining endogeneity in scheduling conflicts using two features of user-paper pairs that, based on prior work, we assume to strongly affect scheduling conflicts. These features are the semantic distance between a user's body of work and the presented paper, and whether a user and the presented paper's author(s) share an affiliation. In section 4.4. we test whether controlling for them affects our estimated treatment effects. We also test the sensitivity of the results to excluding cases with zero scheduling conflicts, on the assumption that if individuals change their schedules to avoid conflicts with particularly important papers, they will reduce conflicts to 0.

We use a linear probability model with the following specification

$$Y_{ij} = \left(\frac{1}{\text{Num. scheduling conflicts}_{ij} + 1}\right) \cdot \tau + P_i + U_j + \epsilon_{ij} \quad (1)$$

where $Y_{ij} = I(citation_{ij} = 1)$ is an indicator for whether paper *i* was cited by user *j* within some time-frame, and $\tau$ measures the effect of scheduling conflicts, which we call the *presentation effect*. $P_i$ is the paper fixed effect, and $U_j$ is the fixed effect for user *j*, and $\epsilon_{ij}$ is the error. We refer to the grand average of the fixed effects $P_i + U_j$ as the *baseline*, capturing all the non-presentation pathways leading to citations, in particular online access. As the denominator ($Num.\,scheduling\,conflicts_{ij} + 1$) goes to infinity, the user has no chance to see any of the presentations and the expression reduces to only the baseline and error. We also estimate logistic models for our main results to test for possible sensitivity to functional form.

We emphasize that the baseline or "control" relative to which the presentation effect is measured is not a complete lack of access to the information, but access to the paper online or through other non-conference channels. This feature of the design mimics the realistic choice facing conference attendees: to attend a paper presentation in person and access it online for any additional details later or to only access it online later.

### 3.5. Descriptive statistics

The 25 computer science conferences comprising the Confer data varied substantially in topic and number of user-paper pairs. Figure 1 below displays the distribution of *Liked* and non-*Liked* user-paper pairs by conference acronym and conference year. Two prominent annual conferences comprise the majority of the data: the ACM Conference on Human Factors in Computing Systems (CHI) and the ACM Conference On Computer-Supported Cooperative Work And Social Computing (CSCW). Topics represented by the other conferences include databases and data management (SIGMOD and VLDB), user interfaces (UIST), and data mining (KDD). Altogether the conferences represent a broad sample of computer science topics.

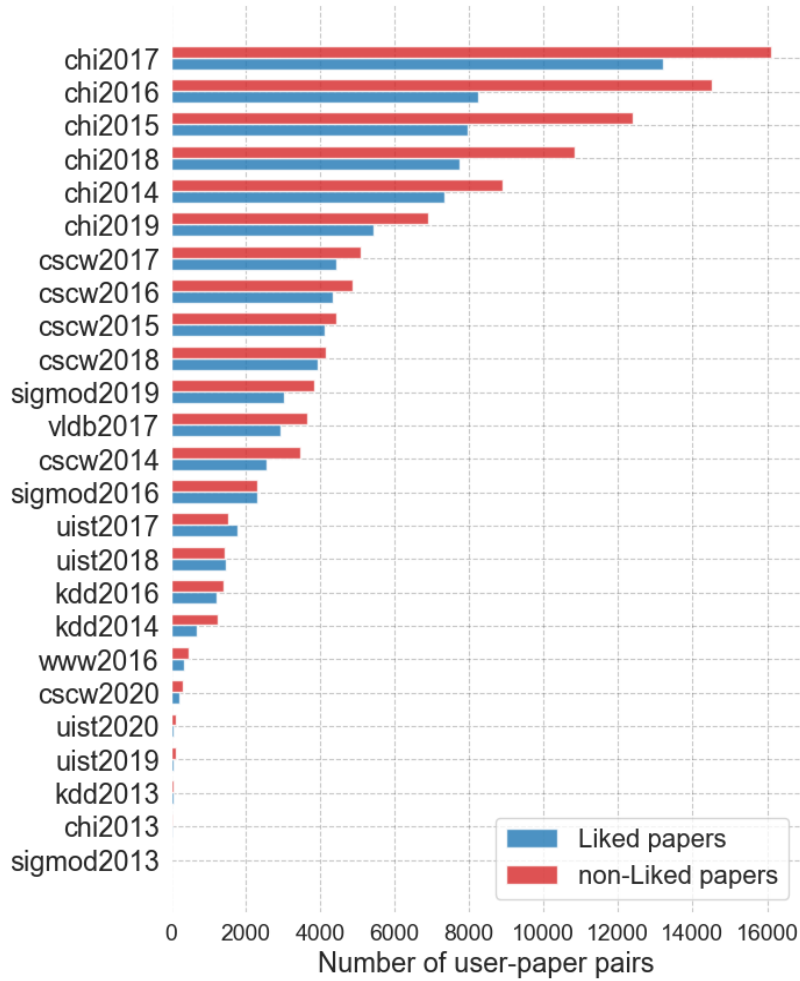

*Figure 1. Distribution of Liked and non-Liked user-paper pairs by conference.*

Crucially, all these conferences were archival, meaning that paper presentations were of final papers and not works-in-progress. This feature of the data removes the possibility that diffusion was affected by mature *maturation* (see Section 2).

The number of user-paper pairs contributed by each user was also uneven. Figure 2 displays the distribution of pairs across users. The maximum number of pairs for a user was 1420, which is omitted from the figure for legibility.

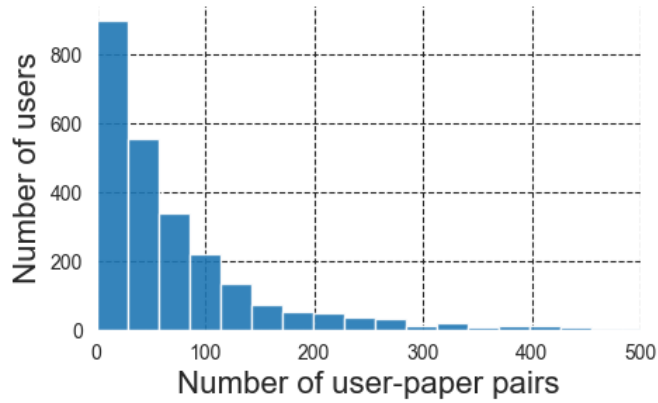

*Figure 2. Histogram of user-paper pairs contributed by each user.*

Figure 3 displays the distribution of scheduling conflicts. The modal user-paper pair for both *Liked* and non-*Liked* papers had no scheduling conflicts. In less than 1% of cases, users had four or more scheduling conflicts in a timeslot.

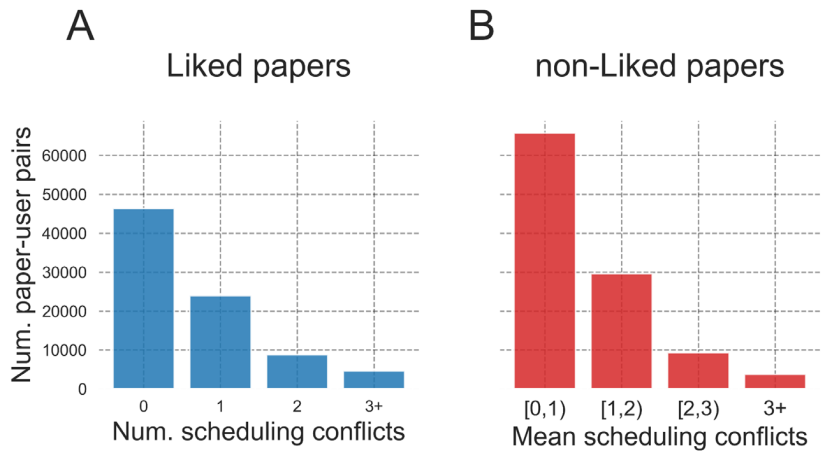

*Figure 3. Distribution of scheduling conflicts for* Liked *(Panel **A**) and non-*Liked *(Panel **B**) user-paper pairs.*

Table 1 shows the descriptive statistics for *Liked* user-paper pairs and Table 2 for non-*Liked* pairs.

|  | count | mean | std | min | 25% | 50% | 75% | max |
|---|---|---|---|---|---|---|---|---|
| **num. scheduling conflicts** | 83726.0 | 0.688 | 0.972 | 0.000 | 0.000 | 0.000 | 1.000 | 11.00 |
| **is cited within 2 yrs** | 83705.0 | 0.035 | 0.184 | 0.000 | 0.000 | 0.000 | 0.000 | 1.00 |
| **is cited within 5 yrs** | 74874.0 | 0.055 | 0.228 | 0.000 | 0.000 | 0.000 | 0.000 | 1.00 |
| **is cited within 9 yrs** | 10725.0 | 0.063 | 0.242 | 0.000 | 0.000 | 0.000 | 0.000 | 1.00 |
| **has shared affiliation** | 83515.0 | 0.028 | 0.164 | 0.000 | 0.000 | 0.000 | 0.000 | 1.00 |
| **semantic distance** | 72397.0 | 0.159 | 0.036 | 0.043 | 0.133 | 0.152 | 0.179 | 0.31 |
| **semantic distance bin** | 72397.0 | 1.374 | 1.141 | 0.000 | 0.000 | 1.000 | 2.000 | 3.00 |

*Table 1. Descriptive statistics for* Liked *papers.*

|  | count | mean | std | min | 25% | 50% | 75% | max |
|---|---|---|---|---|---|---|---|---|
| **num. scheduling conflicts** | 108402.0 | 0.606 | 0.851 | 0.000 | 0.000 | 0.00 | 1.000 | 8.000 |
| **is cited within 2 yrs** | 108369.0 | 0.008 | 0.091 | 0.000 | 0.000 | 0.00 | 0.000 | 1.000 |
| **is cited within 5 yrs** | 95989.0 | 0.016 | 0.124 | 0.000 | 0.000 | 0.00 | 0.000 | 1.000 |
| **is cited within 9 yrs** | 16355.0 | 0.021 | 0.143 | 0.000 | 0.000 | 0.00 | 0.000 | 1.000 |
| **has shared affiliation** | 108134.0 | 0.015 | 0.120 | 0.000 | 0.000 | 0.00 | 0.000 | 1.000 |
| **semantic distance** | 94632.0 | 0.165 | 0.034 | 0.046 | 0.141 | 0.16 | 0.184 | 0.319 |
| **semantic distance bin** | 94632.0 | 1.596 | 1.091 | 0.000 | 1.000 | 2.00 | 3.000 | 3.000 |

*Table 2. Descriptive statistics for serendipity papers.*

The correlation tables for *Liked* and non-*Liked* user-paper pairs are displayed in Figures S4 and S5 in the *SI*.

## 4. Results

### 4.1. The effect of presentations on citations

Within two years of the conference, users cited 3.5% of the papers they *Liked* and 0.8% of the non-*Liked* papers. Figure 4 shows that in the raw data, *i.e.* without any regression adjustments, citation probabilities decreased with the level of scheduling conflicts.

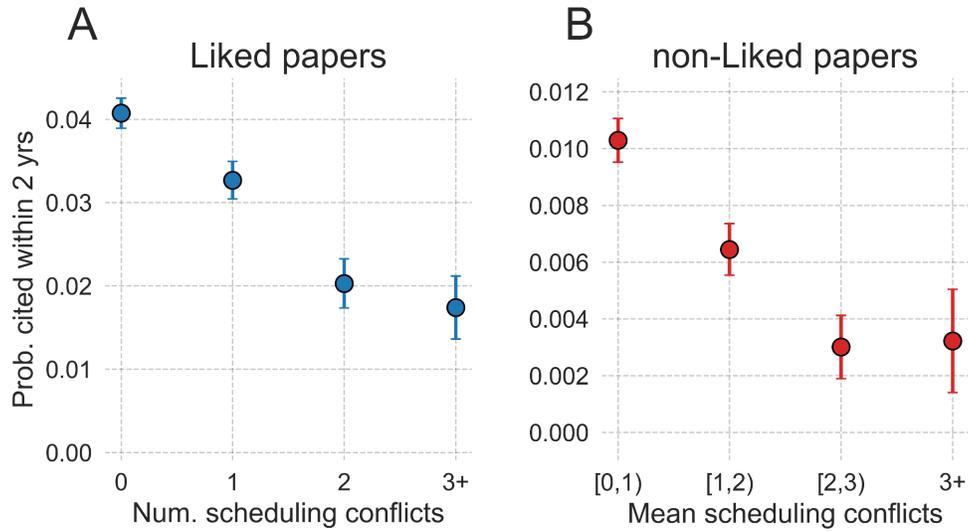

*Figure 4*. Unadjusted (raw data) citation probabilities across user-paper pairs with different levels of scheduling conflicts for **A**: *Liked* user-paper pairs and **B**: non-*Liked* ones. Non-*Liked* scheduling conflicts are the mean of the *Liked* papers' conflicts in that session.

These differences in raw citation probabilities may reflect characteristics beyond just the number of scheduling conflicts. They could be confounded by differences among papers or users. For example, paper presentations with fewer scheduling conflicts might be of higher quality because conference organizers scheduled them in less busy timeslots in the overall schedules. Accordingly, we use fixed effects models with specification (1) and vary the citation time window from one to nine years after the conference. Estimated baseline and presentation effects are displayed in Figure 5, while Table 3 is the regression table for *Liked* papers and Table 4 for non-*Liked* ones. The Tables highlight that the number of observations decreases with the length of the time window; for example, only conferences in 2013 have accrued a nine-year citation time window.

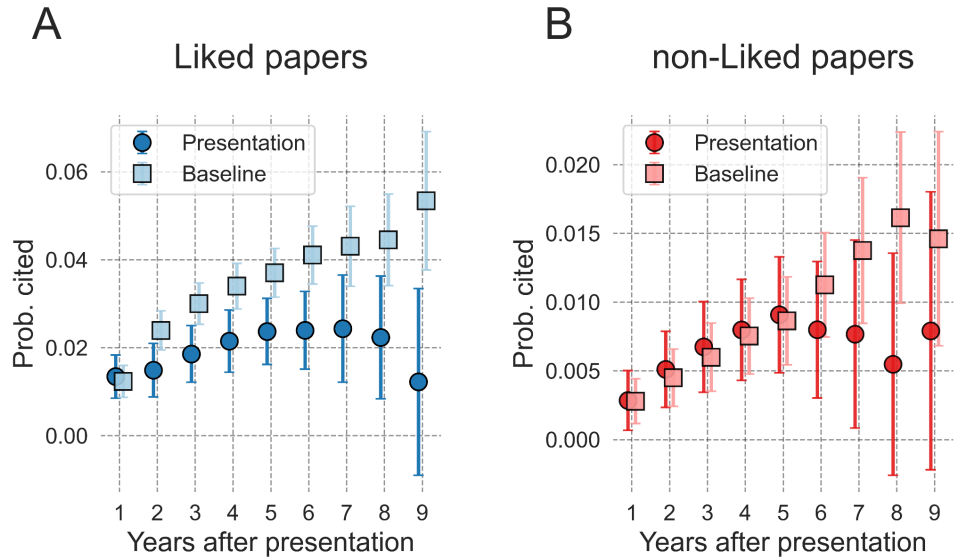

*Figure 5*. Presentation and baseline effects from regressions of form (1) predicting citation 1 to 9 years after the conference. Presentation effects are the coefficients of the 1/(num. scheduling conflicts) term and baseline effects are grand means of the fixed effects. **A**: Liked user-paper pairs and **B**: non-Liked ones. Error bars represent 95% confidence intervals with standard errors clustered at the level of the fixed effects.

Figure 5 Panel A shows that the estimated presentation effects are substantively and statistically significant for almost all years after the conference; the statistically non-significant estimates are for the longest time windows, which have the fewest observations. For example, two years after a conference users cite 2.4% of papers they *Liked* without seeing presentations (baseline), and an extra 1.5% if they could see them presented (presentation effect), which represents a 62.5% boost. For non-*Liked* papers, the analogous estimates are baseline = 0.4%, and presentation effect = 0.5%, which represents a 125% boost. Such a large presentation effect supports the interpretation of non-*Liked* paper citations as serendipitous diffusion, as it was very unlikely to happen without presentations.

We underscore that these effects appear despite attendees having access to all of these papers online. Furthermore, if treatment is defined as seeing rather than having opportunities to see presentations, then our presentation effects are intention-to-treat, and the treatment effects are likely larger. These results answer Research Questions 1 and 2 in the affirmative: in-person presentations increase intentional and serendipitous diffusion, and are especially important for the latter.

|  | \multicolumn{9}{c}{*Dependent variable*: Liked paper is cited within *t* years} |
| --- | --- | --- | --- | --- | --- | --- | --- | --- | --- |
|  | (1) | (2) | (3) | (4) | (5) | (6) | (7) | (8) | (9) |
| Intercept | 0.012*** | 0.024*** | 0.030*** | 0.034*** | 0.037*** | 0.041*** | 0.043*** | 0.044*** | 0.053*** |
|  | (0.002) | (0.002) | (0.002) | (0.003) | (0.003) | (0.003) | (0.005) | (0.005) | (0.008) |
| Presentation effect | 0.013*** | 0.015*** | 0.019*** | 0.021*** | 0.024*** | 0.024*** | 0.024*** | 0.022*** | 0.012 |
|  | (0.003) | (0.003) | (0.003) | (0.004) | (0.004) | (0.005) | (0.006) | (0.007) | (0.011) |
| Observations | 83705 | 83705 | 83705 | 83412 | 74874 | 61707 | 39323 | 22832 | 10725 |
| N. of groups | 2379 | 2379 | 2379 | 2377 | 2175 | 1942 | 1393 | 897 | 506 |
| $R^2$ | 0.001 | 0.000 | 0.001 | 0.001 | 0.001 | 0.001 | 0.001 | 0.001 | 0.000 |
| Residual Std. Error | 0.003 (df=75598) | 0.004 (df=75598) | 0.004 (df=75598) | 0.005 (df=75480) | 0.006 (df=68010) | 0.006 (df=56105) | 0.006 (df=35293) | 0.005 (df=20371) | 0.003 (df=9328) |
| F Statistic | 38.451*** (df=8107; 75598) | 31.137*** (df=8107; 75598) | 39.718*** (df=8107; 75598) | 47.516*** (df=7932; 75480) | 48.029*** (df=6864; 68010) | 37.281*** (df=5602; 56105) | 23.457*** (df=4030; 35293) | 11.760*** (df=2461; 20371) | 1.542 (df=1397; 9328) |

Note: *p<0.1; **p<0.05; ***p<0.01

***Table 3.*** *Fixed effects regression estimates predicting citation of* Liked *papers* t *years after the conference. The model number is* t. *Intercept is the grand mean of the fixed effects. Standard errors are clustered at the level of the fixed effects (user, paper).*

|  | (1) | (2) | (3) | (4) | (5) | (6) | (7) | (8) | (9) |
|---|---|---|---|---|---|---|---|---|---|
| | \multicolumn{9}{c}{*Dependent variable*: non-*Liked* paper is cited within $t$ years} |
| Intercept | 0.003*** | 0.004*** | 0.006*** | 0.008*** | 0.009*** | 0.011*** | 0.014*** | 0.016*** | 0.015*** |
|  | (0.001) | (0.001) | (0.001) | (0.001) | (0.002) | (0.002) | (0.003) | (0.003) | (0.004) |
| Presentation effect | 0.003** | 0.005*** | 0.007*** | 0.008*** | 0.009*** | 0.008*** | 0.008** | 0.005 | 0.008 |
|  | (0.001) | (0.001) | (0.002) | (0.002) | (0.002) | (0.003) | (0.003) | (0.004) | (0.005) |
| Observations | 108369 | 108369 | 108369 | 107861 | 95989 | 78805 | 53482 | 34518 | 16355 |
| N. of groups | 2389 | 2389 | 2389 | 2386 | 2178 | 1955 | 1415 | 924 | 513 |
| $R^2$ | 0.000 | 0.000 | 0.000 | 0.000 | 0.000 | 0.000 | 0.000 | 0.000 | 0.000 |
| Residual Std. Error | 0.001 (df=100117) | 0.001 (df=100117) | 0.002 (df=100117) | 0.002 (df=99787) | 0.002 (df=89022) | 0.002 (df=73092) | 0.002 (df=49333) | 0.001 (df=31923) | 0.002 (df=14845) |
| F Statistic | 9.225*** (df=8252; 100117) | 17.710*** (df=8252; 100117) | 23.423*** (df=8252; 100117) | 26.990*** (df=8074; 99787) | 27.311*** (df=6967; 89022) | 15.149*** (df=5713; 73092) | 8.044*** (df=4149; 49333) | 2.590 (df=2595; 31923) | 2.401 (df=1510; 14845) |

Note: *p<0.1; **p<0.05; ***p<0.01

*Table 4.* Fixed effects regression estimates predicting citation of non-*Liked* papers t *years after the conference. The model number is* t. *Intercept is the grand mean of the fixed effects. Standard errors are clustered at the level of the fixed effects (user, paper).*

## 4.2. Relative contributions of direct and serendipitous diffusion

To compare the magnitude of intentional vs. serendipitous diffusion, we make the following back-of-the-envelope calculation. Focusing on 2-year citations, we assume that the presentation effect for each *Liked* user-paper pair is $\tau_{Liked} = 0.015$ (Table 3, column 2) and for each non-*Liked* one is $\tau_{non-Liked} = 0.005$ (Table 4, column 2). We then divide this effect by each user-paper pair's actual number of scheduling conflicts. Summing across pairs separately for the two types of papers gives the expected amount of intentional and serendipitous diffusion. The sums are $intentional.diffusion = \sum_{i=1}^{num.Likes} \frac{\tau_{Liked}}{1+num.conflicts_i} = 934.9$, and $serendipitous.diffusion = \sum_{i=1}^{num.\ non-Likes} \frac{\tau_{non-Liked}}{1+num.conflicts_i} = 410.3$. The relative contribution of serendipity is thus $\frac{410.3}{934.9\ +\ 410.3} = 30.5\%$. We answer Research Question 3 by showing that, conservatively, about 30% of diffusion created by (opportunities to see) presentations was serendipitous.

## 4.3. Semantic distance

Lastly, we explore how presentation effects vary across papers that are semantically close or distant to the user. To do so, we split the semantic distance variable into quartiles, and interact the treatment variable in specification (1) with the quartile indicators. The resulting estimates are visualized in Figure 6 and presented in Table 5.

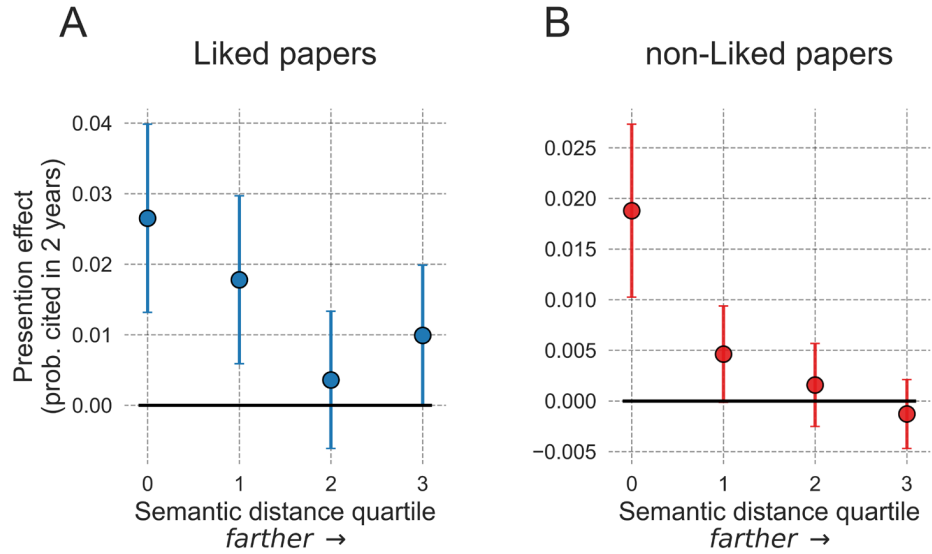

*Figure 6*. Presentation effects (coefficient on scheduling conflicts term in specification (1)) for different semantic distance bins. Dependent variable is citation within two years. **A**: Liked user-paper pairs and **B**: non-Liked ones. Error bars represent 95% confidence intervals with standard errors clustered at the level of the fixed effects.

|  | Dependent variable: Cited within 2 years | | |
|---|---|---|---|
|  | Liked papers | non-Liked papers | non-Liked papers: With mean distance control |
|  | (1) | (2) | (3) |
| Quartile 0 | 0.015* | -0.002 | 0.002 |
|  | (0.008) | (0.004) | (0.003) |
| Quartile 1 | -0.008 | -0.005 | 0.002 |
|  | (0.008) | (0.003) | (0.002) |
| Quartile 2 | -0.013* | -0.006** | 0.002 |
|  | (0.007) | (0.003) | (0.002) |
| Quartile 3 | -0.021*** | -0.008** | 0.002 |
|  | (0.008) | (0.003) | (0.002) |
| Mean distance for liked papers |  |  | 0.005 |
|  |  |  | (0.009) |
| Presentation effect for quartile 0 | 0.027*** | 0.019*** | 0.024*** |
|  | (0.007) | (0.004) | (0.004) |
| Presentation effect for quartile 1 | 0.018*** | 0.005* | 0.008*** |
|  | (0.006) | (0.002) | (0.002) |
| Presentation effect for quartile 2 | 0.004 | 0.002 | 0.005*** |
|  | (0.005) | (0.002) | (0.002) |
| Presentation effect for quartile 3 | 0.010* | -0.001 | 0.001 |
|  | (0.005) | (0.002) | (0.002) |
| Paper fixed effects | Yes | Yes | No |
| User fixed effects | Yes | Yes | No |
| Observations | 83705 | 108369 | 93172 |
| N. of groups | 2379 | 2389 | 2029 |
| $R^2$ | 0.006 | 0.004 | 0.015 |
| Residual Std. Error | 0.014 (df=75591) | 0.005 (df=100110) | 0.012 (df=93163) |
| F Statistic | 55.759*** (df=8114; 75591) | 46.131*** (df=8259; 100110) | 157.199*** (df=9; 93163) |

Note: *p<0.1; **p<0.05; ***p<0.01

*Table 5. Regression estimates predicting citation within two years of the conference as a function of semantic distance between the user's work and the presented paper. Models (1) and (2) are fixed effects models for* Liked *and non-*Liked *papers. Model (3) is an OLS model including a control for the mean semantic distance of* Liked *papers. Standard errors are clustered at the level of the fixed effects.*

The presentation effects decrease with semantic distance and are statistically insignificant at the 0.05 level for the farthest two quartiles. We answer Research Question 4 with the observation that presentations cause diffusion primarily of semantically proximate papers.

There are two plausible interpretations of this pattern. First, presentations have little effect on absorbing semantically distant ideas. Second, users may have not attended distant presentations despite having the opportunities to do so and never got exposed to the ideas. To help adjudicate between these interpretations, we make the assumption that users' choices of which sessions to attend were driven primarily by characteristics of the *Liked* papers, not the non-*Liked* ones. In particular, we assume users were likelier to attend sessions with semantically close *Liked* papers than distant ones. Conditional on this distance to Liked papers, sometimes the non-*Liked* papers in those sessions were distant and sometimes close, resulting in variation in semantic distance that was plausibly unrelated to attendance.

Following these assumptions, we estimate citation rates to non-*Liked* papers after adding a control for the mean semantic distance of *Liked* papers in those sessions. The distance for non-*Liked* papers is highly correlated with the mean distance for *Liked* papers ($\rho$=0.79, p<0.001), leading to inflated standard errors in the fixed effects models[3]. Accordingly, we estimate the model without fixed effects and show the estimates in Table 5, Model 3. Results show that citation rates are much lower for semantically distant non-*Liked* papers even after controlling for *Liked* paper distance, suggesting that these ideas are harder to absorb.

### 4.4. Testing the identification assumption and functional form

We interpret the presentation effect $\tau$ in (1) as causal because we assume scheduling conflicts are exogenous conditional on the fixed effects. How credible is this assumption? We consider two potential sources of endogeneity and test for them with alternative model specifications.

First, users might curate their schedules to avoid scheduling conflicts for presentations they are most interested in seeing. Given the extensive evidence that semantic and institutional proximity predict diffusion (Azoulay et al., 2011; Belenzon & Schankerman, 2013; Catalini, 2017; M. T. Hansen, 1999; Head et al., 2019; van der Wouden & Youn, 2023; Wuestman et al., 2019), we assume that two key proxies of interest are semantic proximity and if the presenter is from the same institution as the user. If accounting for these two proxies does not meaningfully change the results, then it is unlikely that other confounders will do so either. We estimated models of form (1) for two-year citations with and without controls for semantic distance and shared institution and display the results in Table S2 in the *SI*. Strikingly, adding the controls did not change the estimated presentation effect for either *Liked* (models 1 and 2) or non-*Liked* papers (models 4 and 5).

---

[3] These estimates are available upon request.

Second, consider *how* a user might curate their schedule, if they wish to do so[4]. If after *Liking* papers and generating her schedule the user discovers that a presentation of particular interest is conflicting with others, she is likely to delete all the others from the schedule, *i.e.* ensure there are zero scheduling conflicts. In contrast, timeslots with one or more conflicts are unlikely to have been curated in this way. Accordingly, we estimated models after excluding all user-paper pairs with zero scheduling conflicts and display the results in Table S2, models 3 and 6. The exclusion drops by about half the number of observations, and greatly *increases* estimated presentation effects (1.5% → 4.3% for *Liked* papers and 0.5% → 1.8% for non-*Liked*). A possible explanation for the much larger estimates in the subset is that younger scholars *Like* more papers, more often have >1 conflict, are overrepresented in the subset, and are more affected by presentations.

Overall, both tests provide support for the identification assumption that scheduling conflicts are conditionally exogenous.

The analyses throughout the paper relied on linear probability models for interpretive ease, despite the outcomes being binary. To test for the sensitivity of the results to functional form we estimated logistic models predicting citation within two years of the conference. Coefficients and average marginal effects from these models are displayed in Tables S3 and S4, respectively. Many user- and paper- groups had no variation in the outcome variable and were dropped during estimation, resulting in much smaller sample sizes. The coefficient of the presentation effect was positive and statistically significant in all models, matching our preferred specifications. The average marginal presentation effects were generally much larger than in the linear probability models, likely due to the more selected samples used in estimation. We conclude that the results from the linear probability models are consistent with and possibly more conservative than those from logistic models.

## 5. Discussion

This study explored the role in diffusion of seeing ideas presented in person when the same ideas are easily accessible online. The information environment where individuals can access the same ideas through different modalities is an increasingly common but rarely studied one, and it is not a *priori* clear how much value short in-person presentations add to online access, if any. We posed four research questions about presentations and developed answers by combining fine-grained data from *Confer*, a platform researchers use to create personalized schedules at academic conferences, and researchers' subsequent citing decisions. The identification strategy used scheduling conflicts, which we argued are plausibly exogenous after conditioning on user and paper fixed effects. Scheduling conflicts create variation in whether an individual can see a paper presented in-person without changing other determinants of diffusion, such as underlying knowledge or interest. In this research design the treatment is opportunities to see presentations, and we called the associated causal effects "presentation effects." We note that another interpretation of these effects is as intent-to-treat, where the treatment is seeing presentations. Further, we utilized the fact that in *Confer* users can browse all papers and *Like* only those they want added to their schedules. *Liking* provides an explicit measure of intention to learn, enabling us to distinguish between intentional (*Liked*) and serendipitous diffusion (non-*Liked* papers).

---

[4] We thank Nicholas Bloom for this idea.

Research Questions 1 and 2 ask whether in-person presentations increase intentional and serendipitous diffusion. Focusing on citations within two years after the conference, the presentation effects increased intentional diffusion by 62.5% (2.4% → 3.9%) and serendipitous diffusion by 125% (0.4% → 0.9%). Presentations were thus quite effective at stimulating diffusion of both types.

Mechanistically, presentations could have stimulated the diffusion of *Liked* papers by improving engagement and non-*Liked* papers by improving awareness and engagement. Because the presentation effect was, in relative terms, much larger for non-*Liked* papers, presentations likely increased awareness of unsought ideas. That so many evidently useful non-*Liked* papers would not have been cited without presentations also supports our interpretation of this diffusion as serendipitous.

Research Question 3 concerns the contribution of serendipity to diffusion. A back-of-the-envelope calculation showed that even using our narrow definition of serendipitous diffusion as that of non-Liked papers, that contribution was 30.5% of the total diffusion caused by presentations two years post-conference. To our knowledge, this is the first such quantification and it reveals serendipity plays a very substantial role at in-person conferences.

Lastly, Research Question 4 concerns the role of semantic distance in diffusion. We found that presentation effects were precipitously lower for papers that were semantically distant from the researchers and were statistically non-significant at the farthest levels. This pattern leads to two implications. First, it corroborates previous work showing inefficiencies in scientific communication (D. R. Feenberg et al., 2015). If communication was very efficient, researchers would readily utilize ideas that are useful and can be found and absorbed at low cost, and semantically proximate ideas should be relatively easy to find and absorb. Yet even ~15-minute presentations greatly increase the utilization of proximate ideas.

Second, previous work has found that researchers identify semantically distant ideas as particularly influential for their work (Duede et al., 2024), raising the question of whether such diffusion can be stimulated exogenously, for example via events designed to promote interdisciplinarity (Lane et al., 2021). Our results suggest that the effectiveness of such efforts is likely to be very limited.

This study has several limitations, which offer fruitful avenues for future work. First, we use observational data and rely on the identification assumption that scheduling conflicts are conditionally independent of potential outcomes; experimental and quasi-experimental research designs would help establish causality more conclusively. Second, our data do not measure actual presentation attendance, resulting in intent-to-treat effects of presentations. Intent-to-treat effects are particularly useful for policy-makers and administrators who may be interested in encouraging researchers to see ideas that are interdisciplinary or of other types but cannot force them to do so. The effects of actual attendance can be expected to be even larger than our intent-to-treat estimates. Third, we analyzed conferences only in the field of computer science. The role of presentations in other fields is less clear and deserves investigation. On the one hand, presentations in fields where conferences are non-archival may be even more important as they may enable attendees to provide feedback, adding a "reverse" direction of knowledge flow (Leon and McQuillin 2018). On the other hand, attendees may invest less effort into attending presentations of works-in-progress. Lastly, our sample of conferences was in-person, and enabled measuring the added value of presentations when ideas are easily available online. However, it is also important to compare different communication and conferencing modalities directly, which we hope future work will do.

## 6. Conclusion

Knowledge flows between innovators shape the ideas they draw on and affect the rate and direction of their efforts. Two aspects of knowledge flows - geographic proximity and serendipity - continue to be a major focus and debate in the literature. A central challenge of previous work has been to access the micro-level of these flows, how individuals encounter, and learn and utilize specific ideas. Another challenge is establishing causality, since individuals do not invest attention into ideas at random. Lastly, most existing studies do not address changes to the information environment: ideas are increasingly accessible online, so individuals often decide not whether to access ideas or not but whether to access them in ways additional to online.

This paper partially overcomes these challenges by using fine-grained measures of idea diffusion in the setting of academic conferences. We found that in-person presentations increased the diffusion of papers of interest and unsought ones by very large margins despite all the papers being easily available online. Placing these findings within the literature on geographical proximity and innovation leads to three implications. First, the findings suggest that in-person communication does offer advantages that are not completely substitutable by online access to the same information. Those advantages likely lie in increasing awareness of semantically related ideas, as well as engagement with their details. It is important to note the timeframe of our evidence – conferences held between 2013 and 2020. Some recent scholarship has argued that progress in the technologies enabling remote collaboration has materially changed the role of geography in innovation. For example, Presidente and Frey found that research by teams collaborating remotely was less likely to lead to breakthroughs before 2010, after which the pattern reversed (Presidente & Frey, 2022). Relatedly, Head and co-authors find that in mathematics geographical proximity does not predict citations, once interpersonal ties are controlled for, after 2004. It is tempting to hypothesize from such studies that geographical proximity plays little role in diffusion in recent years. Our findings caution against such a sweeping conclusion, as even in recent years seeing ideas presented in-person was important for diffusion.

Second, the effects of proximity are unlikely to be entirely mediated by collaboration ties and the geographic distribution of ideas, which are also geographically concentrated (Head et al., 2019; Wuestman et al., 2019). Third, it is important to note that the large presentation effects occurred through co-location that was very temporary (~15-minute presentations), suggesting as a few others have done that long-term co-location may not be needed to achieve much of its communication benefits (Chai & Freeman, 2019; Lane et al., 2021; Torre, 2008).

We found that about 30% of the diffusion generated by presentations was of ideas that researchers were very unlikely to learn about outside of the conference. If we interpret researchers' screening of papers into their schedules as predictions about which ones are worth investing attention into, the substantial amount of serendipity suggests that the predictions are often wrong, and presentations help correct some of these mistakes. An important question for future research is whether other communication methods, such as online conferences, achieve similar effectiveness in stimulating serendipitous diffusion.

The effectiveness of presentations also provides some support for centering conferences around them, instead of other activities, such as informal networking. The diffusion of ideas through formal presentations is also likely to be more inclusive than if informal activities, which are likely to be less inclusive (Forrester 2021), are made central.

Lastly, the paper contributes to the toolkit of social scientists an attractive identification strategy based on scheduling conflicts. Scheduling conflicts are ubiquitous in the daily lives of individuals and affect which ideas and people they encounter, helping estimate the effect of the latter on important outcomes like diffusion and collaboration.

# Acknowledgments

We thank Chris Rider, DK Kryscynski, Cassidy Sugimoto, JP Ioannidis, Nicholas Bloom, and Douaa Sheet for helpful comments and suggestions. We also thank three anonymous reviewers and an associate editor, whose comments inspired major improvement of an earlier draft. We also thank participants of the following conferences and workshops: University of Michigan Ross School of Business Strategy Brownbag, MIT Economic Sociology Working Group, International Conference on Science of Science and Innovation, ASIS&T annual meeting, and International Network of Analytical Sociology annual meeting, and Stanford METRICS International Forum.

Supplementary Information for

# Intentional and serendipitous diffusion of ideas: Evidence from academic conferences

Misha Teplitskiy, Soya Park, Neil Thompson, David Karger

## Contents

# 1. Illustration of the software

*Confer*, the conference scheduling software used in the study, is available as a webpage (https://confer.csail.mit.edu/) and a phone app. Figures S1-S3 below show screenshots of the webpage version of the software.

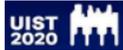

**Fig. S1.** *Confer* main page.

**Fig. S2.** *Confer's* "My Papers" interface and the "Like" functionality.

**Fig. S3**. *Confer's* "My Schedule" interface.

## 2. Full vs. analytic sample

The raw data contained many observations with missing values of critical covariates, particularly presentation type and the timeslots of presentations. Additionally, the conferences contained a number of sessions that are of a social or celebratory nature, such as banquets, keynotes or poster sessions. Consequently, several exclusion criteria were applied to the full sample to reduce it to the "analytic sample" used in the analyses.

Of the 3,061 sessions across all the conferences 2,177 had a valid "type". Of the ones with a valid type 1,150 are classified as "paper", "workshop" or "journal". We sought to exclude the types representing non-research sessions. The excluded types are shown in Table S1, comprising types such as "lunch", "coffee", and "tutorial."

Excluded sessions were not used to track citations but may nevertheless contribute to scheduling conflicts, *i.e.* an attendee might have a conflict between a research paper session and a banquet. We include these non-research sessions when measuring scheduling conflicts. As a robustness check, we also measure conflicts excluding these sessions, and find that the two measures correlated at 0.994. Consequently, we rely on the former measure in the rest of the analyses.

| Keyword | Number of sessions |
|---|---|
| tutorial | 65 |
| course | 60 |
| coffee | 56 |
| student | 42 |
| lunch | 41 |
| keynote | 38 |
| poster | 38 |
| panel | 33 |
| award | 31 |
| demo | 31 |
| reception | 28 |
| registration | 12 |
| lasting impact | 7 |
| banquet | 5 |
| steering committee | 3 |
| townhall | 3 |
| uist closing | 2 |
| <=1 or >=9 or more presentations in the session (and not one of the keywords above) | 12 |
| Total | 470 |

*Table S1*. *The first column contains keywords from titles of sessions that we assumed to indicate a non-research session. The second column shows how many session titles included that keyword and were excluded from the analytic sample.*

Second, we excluded from the analytic sample user-paper pairs where the user had cited the paper before the conference. Such cases occurred in "Lasting Impact" sessions, where the papers presented had been published many years prior to the conference, or the paper was matched to an

*Arxiv* preprint. There were 68 such cases. Other steps in the construction of the analytic sample are described in the main manuscript.

## 3. Supplementary tables

*Table S2. Robustness check – comparing models with and without semantic distance and shared institutions controls and zero conflicts.* Standard errors are clustered at the level of (user, presented paper).

|  | \multicolumn{6}{c}{Dependent variable: Cited within 2 years} |
|---|---|---|---|---|---|---|
|  | Liked papers | Liked papers with controls | Liked papers excl. 0 conflicts | non-Liked papers | non-Liked papers with controls | non-Liked papers excl. zero conflicts |
|  | (1) | (2) | (3) | (4) | (5) | (6) |
| Presentation effect | 0.015*** | 0.015*** | 0.043*** | 0.005*** | 0.005*** | 0.018*** |
|  | (0.003) | (0.003) | (0.012) | (0.001) | (0.002) | (0.004) |
| Shared institution |  | 0.040*** |  |  | 0.010** |  |
|  |  | (0.008) |  |  | (0.004) |  |
| Semantic distance |  | -0.926*** |  |  | -0.297*** |  |
|  |  | (0.067) |  |  | (0.029) |  |
| Paper fixed effects | Yes | Yes | Yes | Yes | Yes | Yes |
| User fixed effects | Yes | Yes | Yes | Yes | Yes | Yes |
| Observations | 83705 | 72377 | 37354 | 108369 | 94593 | 49443 |
| N. of groups | 2379 | 2022 | 1671 | 2389 | 2040 | 1601 |
| $R^2$ | 0.000 | 0.010 | 0.001 | 0.000 | 0.003 | 0.001 |
| Residual Std. Error | 0.004 (df=75598) | 0.018 (df=64779) | 0.004 (df=30382) | 0.001 (df=100117) | 0.005 (df=86858) | 0.002 (df=42349) |
| F Statistic | 31.137*** (df=8107; 75598) | 207.344*** (df=7598; 64779) | 15.249*** (df=6972; 30382) | 17.710*** (df=8252; 100117) | 96.482*** (df=7735; 86858) | 24.211*** (df=7094; 42349) |

Note: *p<0.1; **p<0.05; ***p<0.01

*Table S3. Robustness check – logistic regression.* Regression estimates predicting citation within two years of the conference as a function of the presentation effect (1/scheduling conflicts) and the controls semantic distance and shared institution. Models 1 and 3 for Liked and non-Liked papers use no fixed effects or controls. Models 2 and 4 add fixed effects and Models 3 and 6 add controls. Standard errors are clustered at the level of (user, presented paper) for all models.

| | Liked: simple | Liked: FE | Liked: FE + controls | non-Liked: simple | non-Liked: FE | non-Liked: FE + controls |
|---|---|---|---|---|---|---|
| (Intercept) | -3.914*** | | | -5.667*** | | |
| | (0.100) | | | (0.154) | | |
| Presentation effect | 0.771*** | 0.530*** | 0.510*** | 1.119*** | 1.049*** | 0.912** |
| | (0.101) | (0.116) | (0.124) | (0.172) | (0.247) | (0.278) |
| Semantic distance | | | -40.776*** | | | -60.579*** |
| | | | (2.884) | | | (5.800) |
| Shared institution | | | 0.987*** | | | 1.229*** |
| | | | (0.148) | | | (0.367) |
| Num.Obs. | 83705 | 21545 | 18989 | 108369 | 7158 | 6524 |
| R2 | 0.005 | 0.193 | 0.237 | 0.007 | 0.250 | 0.312 |
| R2 Adj. | 0.005 | -0.111 | -0.075 | 0.007 | -0.186 | -0.131 |
| R2 Within | | 0.003 | 0.056 | | 0.007 | 0.087 |
| R2 Within Adj. | | 0.003 | 0.056 | | 0.006 | 0.086 |
| User fixed effects | No | Yes | Yes | No | Yes | Yes |
| Paper fixed effects | No | Yes | Yes | No | Yes | Yes |

+ p < 0.1, * p < 0.05, ** p < 0.01, *** p < 0.001

*Table S4. Robustness check – average marginal effects. Average marginal effects from logistic regressions predicting citation within two years of the conference as a function of the presentation effect (1/scheduling conflicts) and the controls semantic distance and shared institution. Models 1 and 3 for Liked and non-Liked papers use no fixed effects or controls. Models 2 and 4 add fixed effects and Models 3 and 6 add controls. Standard errors are clustered at the level of (user, presented paper) for all models.*

|  | Liked: simple | Liked: FE | Liked: FE + controls | non-Liked: simple | non-Liked: FE | non-Liked: FE + controls |
|---|---|---|---|---|---|---|
| Presentation effect | 0.026*** | 0.051*** | 0.048** | 0.009*** | 0.089** | 0.074 |
|  | (0.003) | (0.014) | (0.018) | (0.002) | (0.030) | (0.045) |
| Semantic distance |  |  | -3.865*** |  |  | -4.887** |
|  |  |  | (0.580) |  |  | (1.712) |
| Shared institution |  |  | 0.115*** |  |  | 0.128* |
|  |  |  | (0.028) |  |  | (0.064) |
| Num.Obs. | 83705 | 21545 | 18989 | 108369 | 7158 | 6524 |
| R2 | 0.005 | 0.193 | 0.237 | 0.007 | 0.250 | 0.312 |
| R2 Adj. | 0.005 | -0.111 | -0.075 | 0.007 | -0.186 | -0.131 |
| R2 Within |  | 0.003 | 0.056 |  | 0.007 | 0.087 |
| R2 Within Adj. |  | 0.003 | 0.056 |  | 0.006 | 0.086 |
| User fixed effects | No | Yes | Yes | No | Yes | Yes |
| Paper fixed effects | No | Yes | Yes | No | Yes | Yes |

+ p < 0.1, * p < 0.05, ** p < 0.01, *** p < 0.001

# 4. Supplementary figures

***Fig. S4***. *Correlation table among the main variables in the analytic sample for* Liked *user-papers pairs.*

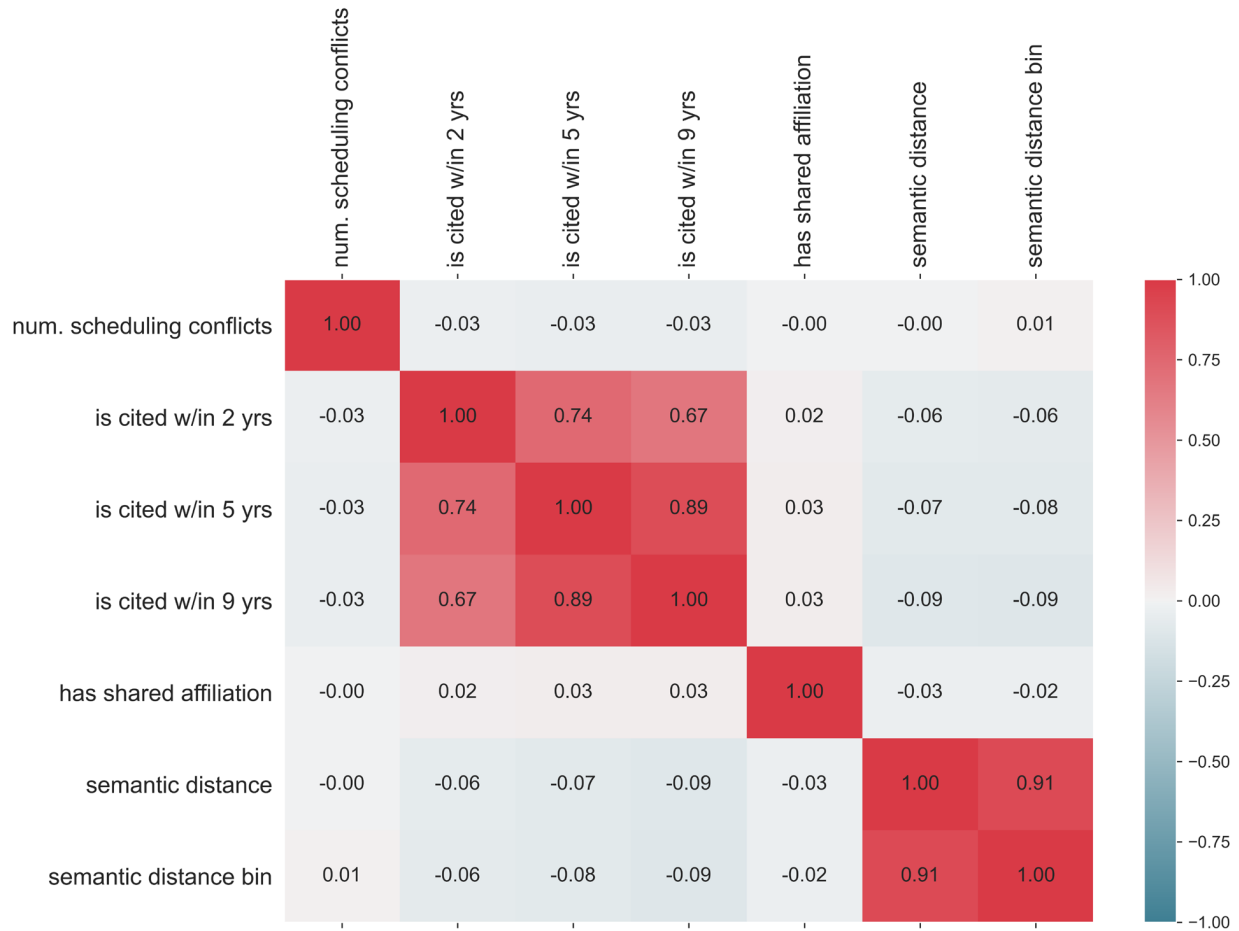

*Fig. S5*. Correlation table among the main variables in the analytic sample for non-Liked *user-papers pairs. challenge*

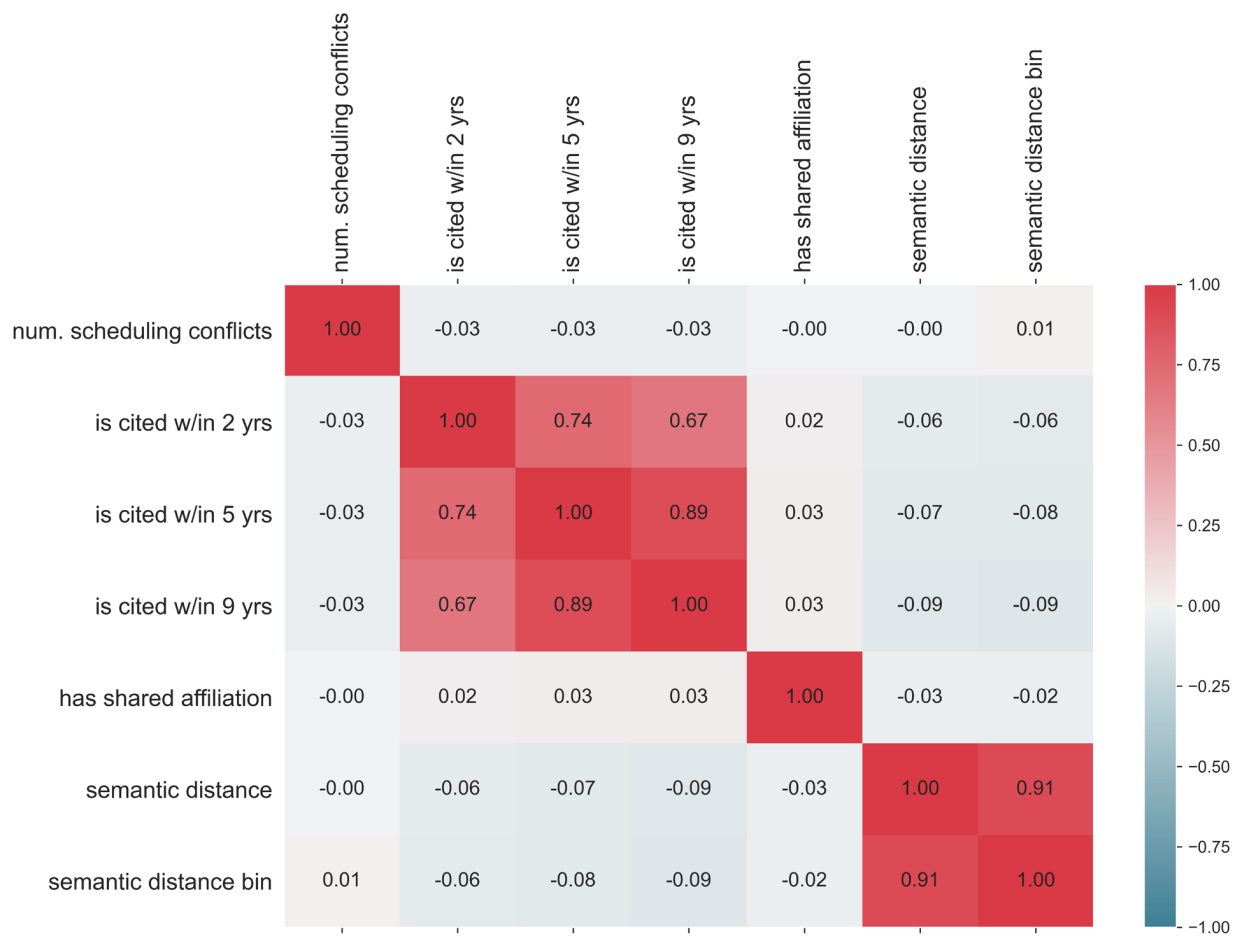